%% file: main.tex
\documentclass[aps,pre,reprint,amsmath,amssymb,superscriptaddress,floatfix]{revtex4-2}

\input{settings}

\begin{document}

\title{Power-Law Injection Spectrum from an Ensemble of \\ Reconnecting Current Sheets}

\author{Omar French}
\homepage{http://www.omarjfrench.com}
\affiliation{Center for Integrated Plasma Studies, Department of Physics, 390 UCB, University of Colorado, Boulder, CO 80309-0390, USA}
\email{omar.french@colorado.edu}

\date{\today}

\begin{abstract}
A collisionless, low-$\beta$ plasma undergoing strong sub-Alfv\'enic turbulence is expected to form a broad hierarchy of reconnecting current sheets immersed in a strong guide magnetic field. If the reconnecting field strength of a sheet scales with its length as a power law of index~$\kappa$, and the number of sheets per unit interval of sheet length decreases as a power law of index~$x$, the current-sheet ensemble injects nonthermal particles with a power-law energy spectrum~$\propto\gamma^{-s}$, where~$s = 1 + (x-2)/(1 + \kappa)$. The spectrum is hard ($s<2$) when the longest sheets deliver most of the injected energy, which requires~$x<3+\kappa$. Past numerical studies have measured~$2\lesssim x\lesssim3.3$, and the scalings of intense current sheets support~$\kappa\gtrsim0$, resulting in~$1\lesssim s\lesssim2.3$. This injection may yield flat or mildly steep radio spectra in optically thin sources and hard ion spectra in the coronae of active galactic nuclei.
\end{abstract}

\maketitle

\input{sections/1_intro}

\input{sections/2_power}

\input{sections/3_applications}

\begin{acknowledgments}
I thank Gregory R. Werner, Maxim Y. Lyutikov, and Stephen Majeski for valuable technical feedback. I also acknowledge support from the National Science Foundation Graduate Research Fellowship under Grant No. DGE 2040434.
\end{acknowledgments}

\section*{Data availability}
No data were created or analyzed in this theoretical article. The model curves in Figs.~\ref{fig:spect} and~\ref{fig:tau-acc} can be reproduced from the equations and parameter values given in the text and captions.

\appendix
\input{sections/4_end}

\bibliography{main}
\bibliographystyle{apsrev4-2}

\end{document}

%% file: settings.tex
\usepackage{graphicx}
\usepackage{dcolumn}

\usepackage[scr=boondoxo]{mathalfa}
\usepackage{hyperref}
\hypersetup{
  colorlinks=true,
  linkcolor=blue,
  citecolor=blue,
  urlcolor=blue
}

%% file: sections/1_intro.tex
\section{Introduction}
A central question in astrophysical plasmas concerns the origin of nonthermal particles \citep{Blandford1987,Petrosian2012}. Recent analytical and numerical studies have shown that relativistic magnetic reconnection in a weak guide magnetic field can produce hard power-law spectra \citep{Sironi2014,Guo2014,Werner2016,Uzdensky2022,HaoZhang2023}, but accessing this regime requires extreme conditions most naturally realized in the vicinity of compact objects, such as pulsars and black holes \citep{Cerutti2012,Hakobyan2022a,Sironi2025}. Although stochastic acceleration can in principle energize particles from the thermal pool \citep{Petrosian2012}, models that apply it across a broad region often require a separate prescription for the suprathermal seed population and its normalization \citep{Fiorillo2024b,Lemoine2025}. Kinetic simulations of magnetized turbulence indeed find that particles are injected out of the thermal pool at reconnecting current sheets \citep{Comisso2018,Comisso2019,Nattila2021,Comisso2022,Guo2025review}. Monte Carlo models of particles crossing ensembles of current sheets find power-law energy distributions whose index depends on the statistics of the sheets \citep{Vlahos2004,Isliker2017}, and a power-law size distribution of plasmoids gives a power-law spectrum \citep{Nishizuka2013}, but a closed-form mapping for current sheets in turbulence has been lacking.

In this article, I show that a hierarchy of reconnecting current sheets embedded in a collisionless, turbulent, low-$\beta$ plasma can supply a nonthermal injection source. Each current sheet considered here operates in the strong guide-field regime and so is not expected to produce an extended hard power-law spectrum by itself \citep{Werner2017a,Werner2021,ComissoJiang2023}, although three-dimensional effects may weaken this expectation \citep{Hoshino2024}. However, if current sheets with lengths in the cascade inertial range are distributed as~$dN_{\rm cs}/d\ell_\perp \propto \ell_\perp^{-x}$, the ensemble can collectively inject suprathermal particles with~$Q_{\rm inj}(\gamma)\propto\gamma^{-s}$ and~$s=1+(x-2)/(1+\kappa)$, where~$\kappa$ is the scaling exponent of the sheet reconnecting field strength with sheet length, $\delta B \propto \ell_\perp^\kappa$. Sheets inheriting the amplitudes of their parent cascade structures have~$\kappa=1/3$ in both dynamically aligned \citep{Boldyrev2006} and Goldreich--Sridhar \citep{GoldreichSridhar1995} turbulence, giving~$s=(3x-2)/4$ if the supply of particles limits the injection of each sheet, while sheets drawn from the most intense structures have~$\kappa$ approaching~$0$~\citep{Chandran2015,MalletSchekochihin2017}, as magnetohydrodynamic (MHD) simulations suggest~\citep{Zhdankin2013,Zhdankin2016}, giving~$s\to x-1$. Past numerical studies also find \(2\lesssim x\lesssim3.3\) \citep{Uritsky2010,Zhdankin2016,Serrano2025}. For~$\kappa=0$, this range corresponds to~$1\lesssim s\lesssim2.3$, and the injected spectrum is hard ($s<2$) for~$x<3$. I then show that the injection is hard when the longest sheets carry most of the injected energy.

If this injected slope survives subsequent source-specific acceleration, cooling, and escape, and if pitch-angle scattering isotropizes the injected electrons, synchrotron emission from electrons accelerated by a current-sheet ensemble could produce a flat or mildly steep radio spectrum. This may be important in optically thin sources such as pulsar wind nebulae~\citep{Gaensler2006,Bucciantini2011}, whose spectra cannot be flattened by the partial synchrotron self-absorption that produces flat spectra at the bases of compact jets \citep{Blandford1979,Kaiser2006}. A soft proton spectrum that produces the neutrinos observed from NGC~1068~\citep{IceCube2022,IceCube2026} must harden at low energies for the proton power to remain below the bolometric luminosity of the active nucleus~\citep{Murase2022,Mbarek2024}. Ions injected by a current-sheet ensemble may supply this hardening \citep{Mbarek2024,French2026AGN}.

Throughout, I consider an arbitrary species of charged particle of mass~$m$, charge~$e$, and number density~$n$, in a plasma with a relativistic enthalpy density~$h$ and large-scale magnetic field~$B_0$. The cyclotron frequency of this species is~$\omega_c \equiv eB_0/(mc)$, and its cold magnetization is~$\sigma_c \equiv B_0^2/(4\pi nmc^2)$. The hot magnetization of the plasma is~$\sigma_h \equiv B_0^2/(4\pi h)$, and its bulk Alfv\'en speed in units of~$c$ is~$\beta_A \equiv [\sigma_h/(1+\sigma_h)]^{1/2}$. Table~\ref{tab:symbols} lists the symbols that recur in this article, along with the equations that define them.

\begin{table}[tp!]
\renewcommand{\arraystretch}{1.05}
\caption{\label{tab:symbols}Recurring symbols and the equations at or near which they are defined. Symbols without an equation number are defined in the introduction. A hat denotes a length in units of~$L_0$. ``Current sheet'' means an actively reconnecting current sheet.}
\begin{ruledtabular}
\begin{tabular}{lll}
Symbol & Meaning & Eq. \\
\hline
\multicolumn{3}{l}{\textit{Fields and plasma}} \\
$n$ & Number density of the species & \\
$h$ & Relativistic enthalpy density of the plasma & \\
$B_0$ & Large-scale guide magnetic field & \\
$\omega_c$ & Cyclotron frequency~$\omega_c \equiv eB_0/(mc)$ & \\
$\sigma_c$ & Cold magnetization~$\sigma_c \equiv B_0^2/(4\pi nmc^2)$ & \\
$\sigma_h$ & Hot magnetization~$\sigma_h \equiv B_0^2/(4\pi h)$ & \\
$\beta_A$ & Bulk Alfv\'en speed over~$c$, $[\sigma_h/(1+\sigma_h)]^{1/2}$ & \\
$\delta B$ & Reconnecting magnetic field of a sheet & \eqref{eq:deltaB} \\
$M_A$ & Alfv\'enic Mach number~$M_A \simeq \delta B(L_0)/B_0$ & \eqref{eq:deltaB} \\
$N_{\rm cs}$ & Number of current sheets in the ensemble & \eqref{eq:cs_dist} \\
$b_g$ & Guide-field strength~$b_g \equiv B_0/\delta B$ & \eqref{eq:guideB} \\
$\delta\beta_A$ & Alfv\'en speed of~$\delta B$ over~$c$, $\beta_A/(b_g^2+\beta_A^2)^{1/2}$ & \eqref{eq:deltabetaA} \\
$\beta_{\rm rec}$ & Reconnection rate~$v_{\rm in}/(c\,\delta\beta_A)\sim0.1$ & \eqref{eq:E_rec_def} \\
$E_{\rm rec}$ & Reconnection electric field~$\beta_{\rm rec}\,\delta B\,\delta\beta_A$ & \eqref{eq:E_rec_def} \\
\hline
\multicolumn{3}{l}{\textit{Scales}} \\
$L_0$ & Outer perpendicular cascade scale &\eqref{eq:deltaB} \\
$\ell_\perp$ & Current sheet length along~$\delta B$ & \eqref{eq:deltaB} \\
$\ell_{\perp,\rm min}$ & Length of the shortest sheet in the ensemble & \eqref{eq:cs_dist} \\
$\ell_{\perp,\rm max}$ & Length of the longest sheet in the ensemble & \eqref{eq:cs_dist} \\
$\ell_\parallel$ & Depth of a sheet of length~$\ell_\perp$, $\ell_\parallel \sim \ell_\perp B_0/\delta B$ & \eqref{eq:ell-parallel-sheet} \\
$\mathscr{d}$ & Sheet half-thickness $\sim$ skin depth & \eqref{eq:ell-parallel-sheet} \\
$\rho_0$ & Nominal gyroradius~$\rho_0 \equiv mc^2/(eB_0)$ & \eqref{eq:delta-gamma-traversal} \\
$A_{\rm cs}$ & Encounter area of a sheet~$A_{\rm cs} \sim2\mathscr{d}\ell_\perp$ & \eqref{eq:A-cs} \\
$\ell_{\perp,\gamma}$ & Shortest accessible sheet~$\max(\ell_{\perp,\rm min},\gamma\rho_0)$ & \eqref{eq:encounter-probability} \\
$V$ & Volume of the turbulent region & \eqref{eq:q-casc} \\
\hline
\multicolumn{3}{l}{\textit{Power-law indices}} \\
$\kappa$ & Sheet-amplitude index, $\delta B\propto\ell_\perp^\kappa$ & \eqref{eq:deltaB} \\
$x$ & Sheet-length index, $dN_{\rm cs}/d\ell_\perp\propto\ell_\perp^{-x}$ & \eqref{eq:cs_dist} \\
$q$ & Energy-jump index, $\Delta\gamma_+\propto\ell_\perp^{q}$ & \eqref{eq:delta-gamma-traversal} \\
$a$ & Encounter-area index, $A_{\rm cs}\propto\ell_\perp^{a}$ & \eqref{eq:A-cs} \\
$s$ & Injection index, $Q_{\rm inj}\propto\chi^{-s}$ & \eqref{eq:s_general_q} \\
\hline
\multicolumn{3}{l}{\textit{Injection}} \\
$\Delta\gamma_\pm$ & Energy jump~$\Delta\gamma_\pm \equiv \pm eE_{\rm rec}\ell_\parallel/mc^2$ & \eqref{eq:delta-gamma-traversal} \\
$\Delta\gamma_0$ & Jump normalization~$\Delta\gamma_0 \equiv \beta_{\rm rec}\beta_A M_A L_0/\rho_0$ & \eqref{eq:delta-gamma-traversal} \\
$\mathcal{I}_k$ & Power-law integral~$\mathcal I_k(y,z) \equiv \int_y^z t^{k-1}\,dt$ & \eqref{eq:Ia_def} \\
$\mathcal{P}_\gamma$ & Sheet-encounter probability density & \eqref{eq:encounter-probability} \\
$\gamma_{\rm th}$ & Lorentz factor of the thermal reservoir & \eqref{eq:inj-kernel} \\
$\chi$ & Lorentz factor above~$\gamma_{\rm th}$, $\chi \equiv \gamma-\gamma_{\rm th}$ & \eqref{eq:inj-kernel} \\
$p_+$ & Positive-work probability per encounter & \eqref{eq:inj-kernel} \\
$N_{\rm th}$ & Number of participating thermal particles & \eqref{eq:inj-kernel} \\
$Q_{\rm inj}$ & One-sided injection source of the ensemble & \eqref{eq:inj-kernel} \\
$P_{\rm casc}$ & Power delivered by the cascade & \eqref{eq:q-casc} \\
$P_{\rm cs}$ & Power of one reconnecting current sheet & \eqref{eq:cs_power} \\
$\epsilon_{\rm inj}$ & Fraction of~$P_{\rm casc}$ used for injection & \eqref{eq:pinj_power_norm} \\
$f_{\rm th}$ & Participating thermal fraction~$N_{\rm th}/(nV)$ & \eqref{eq:fth_max} \\
$f_{\rm th}^{\rm max}$ & Upper bound on~$f_{\rm th}$ & \eqref{eq:fth_max} \\
$\nu_\gamma$ & Encounter rate, neglecting residence times & \eqref{eq:nu-gamma} \\
$\nu_\gamma^{\rm ms}$ & Multi-sheet encounter rate & \eqref{eq:nu-ms} \\
\end{tabular}
\end{ruledtabular}
\end{table}

%% file: sections/2_power.tex
\section{Current-sheet ensemble}
Let us consider a collisionless low-$\beta$ plasma characterized by strong Alfv\'enic turbulence, in which the reconnecting field strength at a current sheet of length~$\ell_\perp$ obeys
\begin{equation} \label{eq:deltaB}
    \delta B(\hat{\ell}_\perp) = M_A \, B_0 \, \hat{\ell}_\perp^\kappa,
    \qquad \, \hat{\ell}_\perp \equiv \ell_\perp/L_0,
\end{equation}
where~$M_A \simeq \delta B(L_0)/B_0$ is the outer-scale Alfv\'enic Mach number. Here~$L_0$ denotes the outer perpendicular scale of the strong Alfv\'enic cascade feeding the current-sheet hierarchy, which need not coincide with the original driving scale in a globally sub-Alfv\'enic system. In a thermally closed plasma, one turnover of this strong cascade transfers an energy density~$\sim M_A^2 B_0^2/8\pi$ to kinetic scales, where it is dissipated and increases the plasma beta by~$\sim M_A^2$. The low-$\beta$ regime considered here therefore requires either~$M_A^2 \ll1$ at~$L_0$, which keeps the beta low only for~$\lesssim M_A^{-2}$ turnovers, or efficient removal of dissipated heat by cooling and/or escape.

\begin{figure}[tp!]
    \centering
    \includegraphics[width=\linewidth]{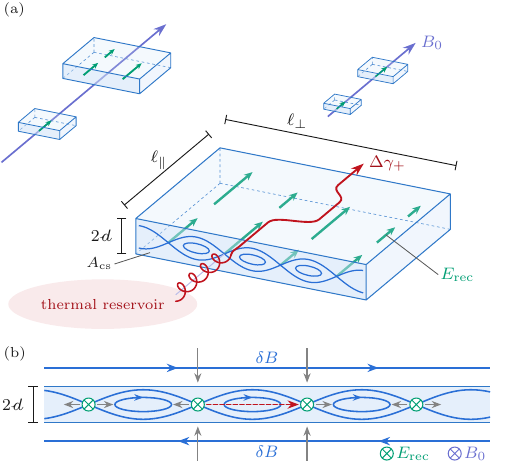}
    \caption{
    Schematic of a reconnecting current-sheet hierarchy, not to scale. (a) Sheets of a range of lengths are threaded by the guide field~$B_0$. A sheet has length~$\ell_\perp$ along the reconnecting field, half-thickness~$\sim \mathscr{d}$ across it, and kink-limited depth~$\ell_\parallel\sim\ell_\perp B_0/\delta B$ along~$B_0$. Guide-field-tied particles from the thermal reservoir enter through thin edges of area~$A_{\rm cs}\sim2\mathscr{d}\ell_\perp$. Green arrows are X-lines, pointing along the reconnection electric field~$E_{\rm rec}$, which is parallel to~$B_0$. They are fragmented along~$B_0$, and a particle streaming through the depth samples several of them before leaving with energy gain~$\Delta\gamma_+$. (b) The thin edge of one sheet, which is its reconnection plane. The upstream field~$\delta B$ spans the full length~$\ell_\perp$, and the layer breaks into a chain of islands separated by X-points that carry~$E_{\rm rec}$. Inflow (gray arrows) carries field lines into the layer, and outflow carries reconnected field lines along it. Following~\citet{Dahlin2017}, I assume that these field lines are stochastic in three dimensions, so that a particle following one returns to the regions near X-lines (dashed track) instead of remaining trapped in an island.
    }
    \label{fig:cs-cartoon}
\end{figure}

Models of intermittent, dynamically aligned Alfv\'enic turbulence contain sheet-like structures across the inertial range \citep{Chandran2015,MalletSchekochihin2017}. I model actively reconnecting sheets as drawn from this population, so that Eq.~\eqref{eq:deltaB} is an assumption about the sheet ensemble. The exponent~$\kappa$ encodes how sheets acquire their reconnecting fields. A sheet may inherit the amplitude of the parent cascade structure from which it forms, as in reconnection-disruption models of Alfv\'enic turbulence \citep{Loureiro2017,Mallet2017MNRAS,Mallet2017JPP}. In dynamically aligned turbulence, the amplitude~$\delta B \propto \lambda^{1/4}$ at field-reversal scale~$\lambda$ \citep{Boldyrev2006} varies as~$\xi^{1/3}$, where~$\xi \propto \lambda^{3/4}$ is the eddy extent along the fluctuation direction~\citep{MalletSchekochihin2017,Loureiro2017} and is identified with the sheet length~$\ell_\perp$ along the reconnecting field. This coincides with the Goldreich--Sridhar scaling~$\delta B \propto \ell_\perp^{1/3}$ \citep{GoldreichSridhar1995}, so inherited amplitudes give~$\kappa=1/3$ for both dynamically aligned and Goldreich--Sridhar turbulence. Relativistic MHD simulations with a strong guide field find the same dynamic alignment~\citep{Chernoglazov2021}, so this value also holds in magnetically dominated plasmas. If active sheets are instead drawn preferentially from the most intense structures, whose amplitudes depend more weakly on scale than those of typical fluctuations \citep{Chandran2015,MalletSchekochihin2017}, the effective exponent decreases, approaching~$\kappa=0$ for scale-independent sheet amplitudes. The value~$\kappa=1/3$ is thus the self-similar scaling of typical fluctuations, while~$\kappa<1/3$ reflects the departure from self-similarity that intermittency produces among the most intense structures. Particles are then energized by magnetic reconnection in the resulting current-sheet ensemble (Fig.~\ref{fig:cs-cartoon}).

Numerical studies find that the lengths of intense current sheets are consistent with a power-law distribution \citep{Zhdankin2016,Serrano2025}
\begin{equation} \label{eq:cs_dist}
    \frac{dN_{\rm cs}}{d\ell_\perp} \propto \ell_\perp^{-x},
    \qquad
    \ell_{\perp,\rm min} \leq \ell_\perp \leq \ell_{\perp,\rm max},
\end{equation}
where~$N_{\rm cs}$ is the number of reconnecting current sheets, the sheet-length index~$x$ sets the relative abundance of small and large sheets (cf.~Fig.~\ref{fig:cs-cartoon}), and~$\ell_{\perp,\rm min}$ and $\ell_{\perp,\rm max}$ are the lengths of the shortest and longest sheets in the ensemble, respectively. Henceforth I will use~$\hat{\ell}_{\perp,\rm min} \equiv \ell_{\perp,\rm min}/L_0$ and~$\hat{\ell}_{\perp,\rm max} \equiv \ell_{\perp,\rm max}/L_0$. The measured indices range from about~$2$~\citep{Uritsky2010,Serrano2025} to~$3.3\pm0.1$~\citep[][Fig.~4 and Table~I]{Zhdankin2016}, and because what sets them is not understood, $x$ will hereafter be kept general. These measurements identify sheets by their current density rather than by whether they reconnect, so applying them here assumes that the reconnecting sheets follow the same length distribution.

Each sheet of length~$\ell_\perp$ carries a reconnecting field of strength~$\delta B(\hat{\ell}_\perp) = M_A \, B_0 \, \hat{\ell}_\perp^\kappa$ [Eq.~\eqref{eq:deltaB}]. In every sheet, let us identify the guide magnetic field (i.e., a coherent non-reversing, out-of-plane magnetic field component) with the large-scale coherent field~$B_0$. The reconnecting field lies perpendicular to~$B_0$, as do the Alfv\'enic fluctuations from which the sheets form. This gives a relative guide-field strength
\begin{equation} \label{eq:guideB}
    b_g(\hat{\ell}_\perp) \equiv \frac{B_0}{\delta B(\hat{\ell}_\perp)} = M_A^{-1} \, \hat{\ell}_\perp^{-\kappa}.
\end{equation}
In what follows, I restrict attention to the part of the current-sheet hierarchy that both lies within the scale range over which Eq.~\eqref{eq:deltaB} applies (i.e., $\ell_\perp \lesssim L_0$) and remains in the strong guide-field regime (i.e., $M_A \hat{\ell}_{\perp, \rm max}^\kappa \ll 1$). Larger sheets that violate either or both of these conditions would require a separate treatment beyond the scope of this article.

Active reconnecting sheets in this ensemble are modeled as ribbons with half-thickness~$\sim \mathscr{d}$ across the reconnecting field, where~$\mathscr{d}$ denotes the kinetic scale of the species, of order its skin depth. Along the guide field, each sheet has a depth~$\ell_\parallel$. Since the plasma is low-$\beta$, pressure-driven instabilities (e.g., those governed by Suydam's criterion \citep{Suydam1958,Goedbloed1973}) are expected to be subdominant to current-driven ones. I therefore take the depth to be limited by the flux-rope kink-instability threshold (i.e., the Kruskal--Shafranov limit \citep{Kruskal1954,Shafranov1956,KruskalTuck1958,Oz2011,Qile2021,Qile2024c}),
\begin{equation} \label{eq:ell-parallel-sheet}
    \ell_\parallel \sim \ell_\perp \, b_g(\hat{\ell}_\perp) = L_0 \, M_A^{-1} \, \hat{\ell}_\perp^{1-\kappa}.
\end{equation}
For amplitudes inherited from the parent cascade, this kink-limited depth scales identically to the critical-balance parallel coherence length of the parent structure, $\sim \ell_\perp b_g(\hat{\ell}_\perp)$, in both dynamically aligned and Goldreich--Sridhar turbulence. The same holds at any~$\kappa$ for intense current sheets in MHD simulations of turbulence with a strong guide field, because their lifetimes follow critical balance~\citep{Zhdankin2015,Zhdankin2016}. Equation~\eqref{eq:ell-parallel-sheet} is thus consistent with the parallel extent available to a sheet at birth. The sheets considered here therefore have lengths and depths in the inertial range and thicknesses at the kinetic scale, where the diffusion region supports the reconnection electric field.

Two measured scalings of these intense current sheets~\citep{Zhdankin2013,Zhdankin2016} constrain~$\kappa$. Their extents along~$B_0$ grow as~$\ell_\perp^{0.9\text{--}1.0}$, which Eq.~\eqref{eq:ell-parallel-sheet} converts to~$\kappa\approx0$--$0.1$. Their thicknesses remain at the dissipation scale, so their Ohmic dissipation rates scale as~$\delta B^2$ times their area, and the measured rates give~$\kappa\approx0.05$--$0.15$. Both estimates carry uncertainties of about~$0.1$. I therefore adopt scale-independent sheet amplitudes ($\kappa=0$) as the fiducial case. I treat~$0 \leq \kappa \leq 1/3$ as the plausible range, because the current-density threshold that selects these sheets may bias both estimates toward~$\kappa=0$ and because the simulations are resistive rather than kinetic. The mapping derived below holds for any~$\kappa$, and applies locally with~$\kappa$ replaced by~$d\ln\delta B/d\ln\ell_\perp$ if the effective exponent varies with scale.

\section{Injection source}
With the current-sheet ensemble defined, let us now compute the one-sided injection source it produces.

\subsection{Energy gain per encounter}
Since each current sheet possesses a strong guide magnetic field [Eq.~\eqref{eq:guideB}], acceleration mechanisms outside diffusion regions, such as Fermi reflections off freshly reconnected field lines \citep{Drake2006a}, are suppressed~\citep{Dahlin2017,French2023,French2026}. Meanwhile, a strong guide field makes the parallel electric field near X-lines the dominant energization channel~\citep{Dahlin2014,Dahlin2017,Sironi2022,French2023,Totorica2023}. I therefore model charged-particle energization as work done by the reconnection electric field, which depends on the reconnecting field~$\delta B$ and its Alfv\'en speed. The latter, normalized to the speed of light, may be written as~\citep{Werner2017a,Uzdensky2022}
\begin{equation} \label{eq:deltabetaA}
    \delta \beta_A(\hat{\ell}_\perp) = \sqrt{ \frac{\delta B^2(\hat{\ell}_\perp)}{\delta B^2(\hat{\ell}_\perp)+B_0^2+4\pi h}}.
\end{equation}
Inserting Eq.~\eqref{eq:guideB} into Eq.~\eqref{eq:deltabetaA} yields
\begin{equation} \label{eq:delta_betaA_full}
    \delta\beta_A(\hat{\ell}_\perp) = \frac{\beta_A}{\sqrt{b_g^2(\hat{\ell}_\perp)+\beta_A^2}}.
\end{equation}
Because~$\beta_A<1$, the strong guide-field condition~$b_g\gg1$ ensures~$b_g\gg\beta_A$, so that~$\delta\beta_A(\hat{\ell}_\perp)\simeq\beta_A/b_g(\hat{\ell}_\perp)=M_A\,\beta_A\,\hat{\ell}_\perp^\kappa\ll1$ for any~$\sigma_h$. The reconnection outflow carries the guide field, whose energy adds to the inertia of the outflow. The outflow therefore remains nonrelativistic even where~$\sigma_h\gg1$, as found by kinetic simulations of \citet{Liu2014}. Relativistic outflows would require~$b_g\lesssim\beta_A$ and are outside the regime considered in this model. The reconnection electric field is then
\begin{equation} \label{eq:E_rec_def}
    E_{\rm rec}(\hat{\ell}_\perp) = \beta_{\rm rec} \, \delta B(\hat{\ell}_\perp) \, \delta \beta_A(\hat{\ell}_\perp),
\end{equation}
where~$\beta_{\rm rec}\equiv v_{\rm in}/(c\,\delta\beta_A)\sim0.1$ is the dimensionless reconnection rate in collisionless plasmas \citep{Liu2017,Haggerty2017,Goodbred2022,Liu2025}, with~$v_{\rm in}$ the speed of the inflow into the sheet, so that~$E_{\rm rec}=v_{\rm in}\,\delta B/c$. Near each X-line, the outflow opens out and bends the upstream field lines, whose tension weakens the field reaching the X-line and limits this rate to~$\sim0.1$ for antiparallel fields~\citep{Liu2017,Liu2025}. Simulations with guide fields up to~$80$ times the reconnecting field find the same rate, independent of system size~\citep{Liu2014,TenBarge2014,Stanier2015a,Stanier2015b}. In the strong guide-field limit, $\delta\beta_A\simeq\beta_A/b_g$, and the reconnection electric field becomes
\begin{equation} \label{eq:E_rec}
    E_{\rm rec}(\hat{\ell}_\perp) = \beta_{\rm rec} \, \beta_A \, B_0 \, M_A^2 \, \hat{\ell}_\perp^{2\kappa}.
\end{equation}
The turbulence enters Eq.~\eqref{eq:E_rec}, through the factor~$M_A^2\hat{\ell}_\perp^{2\kappa}$, only by supplying each sheet with its reconnecting field~$\delta B$. The perpendicular motional electric field of the turbulence energizes particles by stochastic acceleration after their injection~\citep{Comisso2018,Comisso2019} and is not included here.

Let us assume that the effective line-integrated potential drop along the sheet depth scales as~$E_{\rm rec} \ell_\parallel$, with~$E_{\rm rec}$ given by Eq.~\eqref{eq:E_rec} and~$\ell_\parallel$ by Eq.~\eqref{eq:ell-parallel-sheet}. Because~$E_{\rm rec}$ is concentrated at X-lines fragmented along the sheet and along~$B_0$ (Fig.~\ref{fig:cs-cartoon}), this scaling assumes that the stochastic field lines of three-dimensional reconnection return a particle streaming through the depth to the X-lines. \citet{Dahlin2017} found this return for guide fields comparable to the reconnecting field, whereas~$b_g\gg1$ here. Under this assumption, a single sheet traversal imparts to particles the energy increment
\begin{equation} \label{eq:delta-gamma-traversal}
    \Delta \gamma_\pm(\hat{\ell}_\perp) \equiv \pm \frac{eE_{\rm rec}(\hat{\ell}_\perp)\,\ell_\parallel}{mc^2} = \pm \beta_{\rm rec} \, \beta_A \, M_A \, \frac{L_0}{\rho_0} \, \hat{\ell}_\perp^{1 + \kappa},
\end{equation}
where~$\rho_0 \equiv mc^2/(eB_0) = c/\omega_c$ is the nominal gyroradius of the injected species. The sign in Eq.~\eqref{eq:delta-gamma-traversal} reflects that individual sheet traversals may either energize or de-energize particles. Here and below, scale-independent factors of order unity accounting for the finite volume filling fraction and patchy coherence of the reconnection electric field are absorbed into the normalization of the energy jump. It is useful to write Eq.~\eqref{eq:delta-gamma-traversal} as~$\Delta \gamma_\pm = \pm\Delta \gamma_0 \, \hat{\ell}_\perp^q$, where~$\Delta\gamma_0 \equiv \beta_{\rm rec}\beta_A M_A L_0/\rho_0$. In the strong guide-field case considered here, where Eqs.~\eqref{eq:ell-parallel-sheet} and \eqref{eq:E_rec} are valid, $q=1+\kappa$.

\subsection{Injection spectrum}
Let us now calculate the probability that a given particle encounters a sheet of length~\(\ell_\perp\). In general, particles enter current sheets through a characteristic cross-sectional area~$A_{\rm cs}$ that depends on the sheet length~$\ell_\perp$. Suppose the scale-dependent part of~$A_{\rm cs}$ scales as
\begin{equation} \label{eq:A-cs}
    A_{\rm cs}\propto \ell_\perp^a .
\end{equation}
Guide-field-tied particles stream along~$B_0$ and enter a sheet through its thin edge, so their encounter rate is their flux along~$B_0$ through the projected area of this edge. For the ribbon geometry of Fig.~\ref{fig:cs-cartoon}, this area is~$A_{\rm cs} \sim 2\mathscr{d}\ell_\perp$, which is proportional to~$\ell_\perp$ because the thickness is fixed, so $a=1$. If the guide field is instead comparable to the reconnecting field ($b_g\sim1$), particles are no longer guided along~$B_0$ and can arrive at a sheet from a broad range of directions. They then enter mainly through its broad faces, of area~$\sim\ell_\perp\ell_\parallel\propto\ell_\perp^{2-\kappa}$ [Eq.~\eqref{eq:ell-parallel-sheet}], so that~$a\simeq2-\kappa$. Additionally, let us assume that particles can only interact with sheets of length exceeding their gyroradius, which is~$\sim\gamma\rho_0$ once their pitch angles have been isotropized. Using the shorthand
\begin{equation} \label{eq:Ia_def}
    \mathcal I_k(y,z) \equiv \int_y^z t^{k-1} \,dt = \frac{z^k-y^k}{k}, \, \, \mathcal I_0(y,z) = \ln(z/y),
\end{equation}
the probability density of encountering a sheet of length~$\ell_\perp$ is then
\begin{equation}
\begin{split} \label{eq:encounter-probability}
    \mathcal P_\gamma\,d\hat{\ell}_\perp
    &\equiv
    \frac{A_{\rm cs} \, (dN_{\rm cs}/d\hat{\ell}_\perp)}{\int_{\hat{\ell}_{\perp,\gamma}}^{\hat{\ell}_{\perp,\rm max}} A_{\rm cs} \, (dN_{\rm cs}/d\hat{\ell}_\perp)\,d\hat{\ell}_\perp} \,d\hat{\ell}_\perp \\
    &=
    \frac{\hat{\ell}_\perp^{a-x}}{\mathcal{I}_{a+1-x}(\hat{\ell}_{\perp,\gamma},\hat{\ell}_{\perp,\rm max})} \, d\hat{\ell}_\perp,
\end{split}
\end{equation}
where~$\ell_{\perp, \gamma} \equiv \max( \ell_{\perp,\rm min},\,\gamma\rho_0)$ is the length of the shortest sheet accessible to a particle of Lorentz factor~$\gamma$ and~$\hat{\ell}_{\perp, \gamma} \equiv \ell_{\perp, \gamma}/L_0$.

With the energy gain per encounter of Eq.~\eqref{eq:delta-gamma-traversal} and the sheet-encounter probability of Eq.~\eqref{eq:encounter-probability}, I now calculate the injection rate of particles originating from a thermal reservoir. For simplicity, let us approximate the thermal particles that take part in sheet encounters by \(N_{\rm th}\) particles at a representative Lorentz factor \(\gamma_{\rm th}\). Particles undergoing negative-work encounters are returned to the reservoir, whereas particles undergoing positive-work encounters gain energy~$\Delta\gamma_+$. Those gaining~$\Delta\gamma_+ \gtrsim \gamma_{\rm th}$ separate from the thermal population and are injected into the nonthermal population, whereas those with smaller gains remain within the thermal distribution. With~$\chi\equiv\gamma-\gamma_{\rm th}$, the one-sided injection source is
\begin{equation}
\begin{split} \label{eq:inj-kernel}
    Q_{\rm inj}(\chi) = p_+ \, N_{\rm th} \, \nu_{\gamma_{\rm th}}^{\rm ms} \int_{\hat{\ell}_{\perp, \gamma_{\rm th}}}^{\hat{\ell}_{\perp,\rm max}} \mathcal P_{\gamma_{\rm th}} \delta\!\left(\chi-\Delta\gamma_+(\hat{\ell}_\perp)\right)\, d\hat{\ell}_\perp,
\end{split}
\end{equation}
where \(p_+\) is the probability that an encounter performs positive work, and \(\nu_{\gamma_{\rm th}}^{\rm ms}\) is the current-sheet encounter rate of particles in the thermal reservoir, including their residence times in the sheets [see Eq.~\eqref{eq:nu-ms}]. Here, $Q_{\rm inj}$ is a one-sided upward source supplied by a thermal reservoir, not a steady-state particle distribution by itself. It counts every positive-work encounter, including gains below~$\gamma_{\rm th}$. It describes the newly injected population on timescales of order a few~$(\nu_{\gamma_{\rm th}}^{\rm ms})^{-1}$, before multi-sheet drift and diffusion reshape it. Without such a reservoir, the signed jumps should instead be evolved with the drift and diffusion coefficients given in the Appendix, in which net multi-sheet drift of a particle requires~$p_+>1/2$.

Because~$\Delta\gamma_+=\Delta\gamma_0 \hat{\ell}_\perp^q$ increases monotonically with sheet length, each energy gain~$\chi$ corresponds to the single length~$\hat{\ell}_\perp=(\chi/\Delta\gamma_0)^{1/q}$, and Eq.~\eqref{eq:inj-kernel} evaluates to
\begin{equation} \label{eq:Qinj}
    Q_{\rm inj}(\chi)
    =
    \frac{p_+ \, N_{\rm th} \, \nu_{\gamma_{\rm th}}^{\rm ms}/q}{\mathcal{I}_{a+1-x}(\hat{\ell}_{\perp, \gamma_{\rm th}},\hat{\ell}_{\perp,\rm max})}
    \left( \Delta\gamma_0 \right)^{s-1}
    \chi^{-s},
\end{equation}
where
\begin{equation} \label{eq:s_general_q}
    s = 1 + \frac{x-a-1}{q}.
\end{equation}
In the strong guide-field limit considered above, $a=1$ and~$q=1+\kappa$, so
\begin{equation}
    s = 1 + \frac{x-2}{1+\kappa}.
\end{equation}
Thus $s=x-1$ for the fiducial scale-independent sheet amplitudes ($\kappa=0$). For~$\kappa>0$, the mapping assumes that the supply of particles limits the injection of each sheet, as discussed after Eq.~\eqref{eq:cs_power}. It then gives~$s=(3x-2)/4$ for amplitudes inherited from the parent cascade ($\kappa=1/3$). The source in Eq.~\eqref{eq:Qinj} is supported on~$\chi \in [\Delta\gamma_+(\hat{\ell}_{\perp,\gamma_{\rm th}}),\, \Delta\gamma_+(\hat{\ell}_{\perp,\rm max})]$, the range over which the geometric and acceleration scalings above are valid, and vanishes outside this interval (provided~$\hat{\ell}_{\perp,\gamma_{\rm th}}<\hat{\ell}_{\perp,\rm max}$). This interval spans a factor~$(\hat{\ell}_{\perp,\rm max}/\hat{\ell}_{\perp,\gamma_{\rm th}})^{q}$ in energy, and a break in the sheet-length distribution at~$\ell_\perp$ produces a break in~$Q_{\rm inj}$ at~$\chi=\Delta\gamma_+(\hat{\ell}_\perp)$. The injected part of the source, $\chi\gtrsim\gamma_{\rm th}$, comes from sheets longer than~$(\gamma_{\rm th}/\Delta\gamma_0)^{1/q}L_0$, which is~$\approx160\mathscr{d}$ for the fiducials of Fig.~\ref{fig:spect}.

The prediction of Eq.~\eqref{eq:s_general_q} may be tested in kinetic simulations by independently measuring the index~$x$ of the current-sheet length distribution \(dN_{\rm cs}/d\ell_\perp\) (e.g., \citep{Zhdankin2016,Serrano2025,Davis2026}) and the spectral index~$s$ of the distribution of newly injected particles. The scalings~$A_{\rm cs} \propto \ell_\perp^{a}$ and~$\Delta\gamma_+ \propto \ell_\perp^{q}$ may also be directly tested with tracer-particle diagnostics.

\subsection{Normalization}
To bound the amplitude of the injection source, I obtain the dimensionless normalization~$\mathcal{N}_{\rm cs}$ of the sheet distribution~$dN_{\rm cs}/d\hat{\ell}_\perp = \mathcal N_{\rm cs} \, \hat{\ell}_\perp^{-x}$ (equivalently~$dN_{\rm cs}/d\ell_\perp = \mathcal N_{\rm cs} \, L_0^{x-1} \, \ell_\perp^{-x}$) by assuming that the current-sheet ensemble dissipates approximately all of the magnetic energy arriving from the turbulent cascade. With~$L_0$ the outer perpendicular scale of the strong Alfv\'enic cascade that feeds the sheet hierarchy, the magnetic energy density of the outer-scale fluctuations is~$\sim M_A^2 B_0^2/8\pi$ and the turnover time is~$\sim L_0/\left( M_A \beta_A c \right)$, giving, in a turbulent region of volume~$V$,
\begin{equation} \label{eq:q-casc}
    P_{\rm casc} \sim V \frac{B_0^2}{8\pi}\frac{c\beta_A}{L_0} M_A^3 = \int_{\hat{\ell}_{\perp,\rm min}}^{\hat{\ell}_{\perp,\rm max}} P_{\rm cs} \frac{dN_{\rm cs}}{d\hat{\ell}_\perp}\,d\hat{\ell}_\perp,
\end{equation}
where~$P_{\rm cs}$ is the power of a single current sheet of length~$\ell_\perp$, given by
\begin{equation} \label{eq:cs_power}
\begin{split}
    P_{\rm cs} &\sim \frac{\delta B^2(\hat{\ell}_\perp)}{8\pi} \, c \, \delta \beta_A(\hat{\ell}_\perp)\,\beta_{\rm rec}\,2 \ell_\perp \ell_\parallel \\ &\simeq 2 \beta_{\rm rec} \, \frac{B_0^2}{8\pi} \, c \beta_A \, M_A^2 \, L_0^2 \, \hat{\ell}_\perp^{2+2\kappa}.
\end{split}
\end{equation}
Equation~\eqref{eq:cs_power} counts the energy of the reconnecting field carried in through both broad faces of the sheet, which is about half of the incoming Poynting flux. Because each injected particle gains~$eE_{\rm rec}\ell_\parallel$, Eq.~\eqref{eq:cs_power} also caps the injection rate of each sheet at~$P_{\rm cs}/(eE_{\rm rec}\ell_\parallel)\propto\ell_\perp^{1+\kappa}$, and sheets held to this cap give~$s=(x-1)/(1+\kappa)$, which equals the supply-limited~$s=x-1$ at~$\kappa=0$. The normalization resulting from Eq.~\eqref{eq:q-casc} is
\begin{equation} \label{eq:sheet-pop-normalization}
    \mathcal{N}_{\rm cs} = \frac{M_A \, V/L_0^3}{2\beta_{\rm rec} \mathcal{I}_{3+2\kappa-x}(\hat{\ell}_{\perp,\rm min},\hat{\ell}_{\perp,\rm max})}.
\end{equation}
The cascade power~$P_{\rm casc}$ counts only the magnetic fluctuation energy. Order-unity factors in~$P_{\rm casc}$ and~$P_{\rm cs}$ rescale~$\mathcal{N}_{\rm cs}$ and the encounter rates but not the injection index~$s$. The same power budget requires the one-sided injection source to satisfy
\begin{equation} \label{eq:pinj_power_norm}
    mc^{2} \int_{\Delta \gamma_+(\hat{\ell}_{\perp, \gamma_{\rm th}})}^{\Delta \gamma_+(\hat{\ell}_{\perp,\rm max})}
    \chi \, Q_{\rm inj}(\chi) \, d\chi \leq \epsilon_{\rm inj} P_{\rm casc},
\end{equation}
where~$\epsilon_{\rm inj} \leq 1$ is the fraction of cascade power devoted to injection from the thermal reservoir. Using~$\int Q_{\rm inj}(\chi)\,d\chi= p_+ N_{\rm th} \nu_{\gamma_{\rm th}}^{\rm ms}$, the participating thermal fraction~$f_{\rm th} \equiv N_{\rm th}/(nV)$ is at most~$f_{\rm th}^{\rm max}$, where
\begin{equation} \label{eq:fth_max}
    f_{\rm th}^{\rm max} \equiv \frac{M_A \, \epsilon_{\rm inj} \, \sqrt{\sigma_c}}{2p_+} \frac{\mathcal{I}_{3+2\kappa-x}(\hat{\ell}_{\perp,\rm min},\hat{\ell}_{\perp,\rm max})}{\mathcal{I}_{3+\kappa-x}(\hat{\ell}_{\perp,\gamma_{\rm th}},\hat{\ell}_{\perp,\rm max})} \frac{\nu_{\gamma_{\rm th}}}{\nu_{\gamma_{\rm th}}^{\rm ms}},
\end{equation}
where~I take~$\mathscr{d}$ equal to the skin depth, so that~$\mathscr{d}^2/\rho_0^2=\sigma_c$. The final factor, the ratio of the encounter rates of Eqs.~\eqref{eq:nu-gamma} and~\eqref{eq:nu-ms}, accounts for finite sheet residence times and approaches unity when residence times are negligible.

\subsection{Hard and soft injection}
\begin{figure}[tp!]
    \centering
    \includegraphics[width=\linewidth]{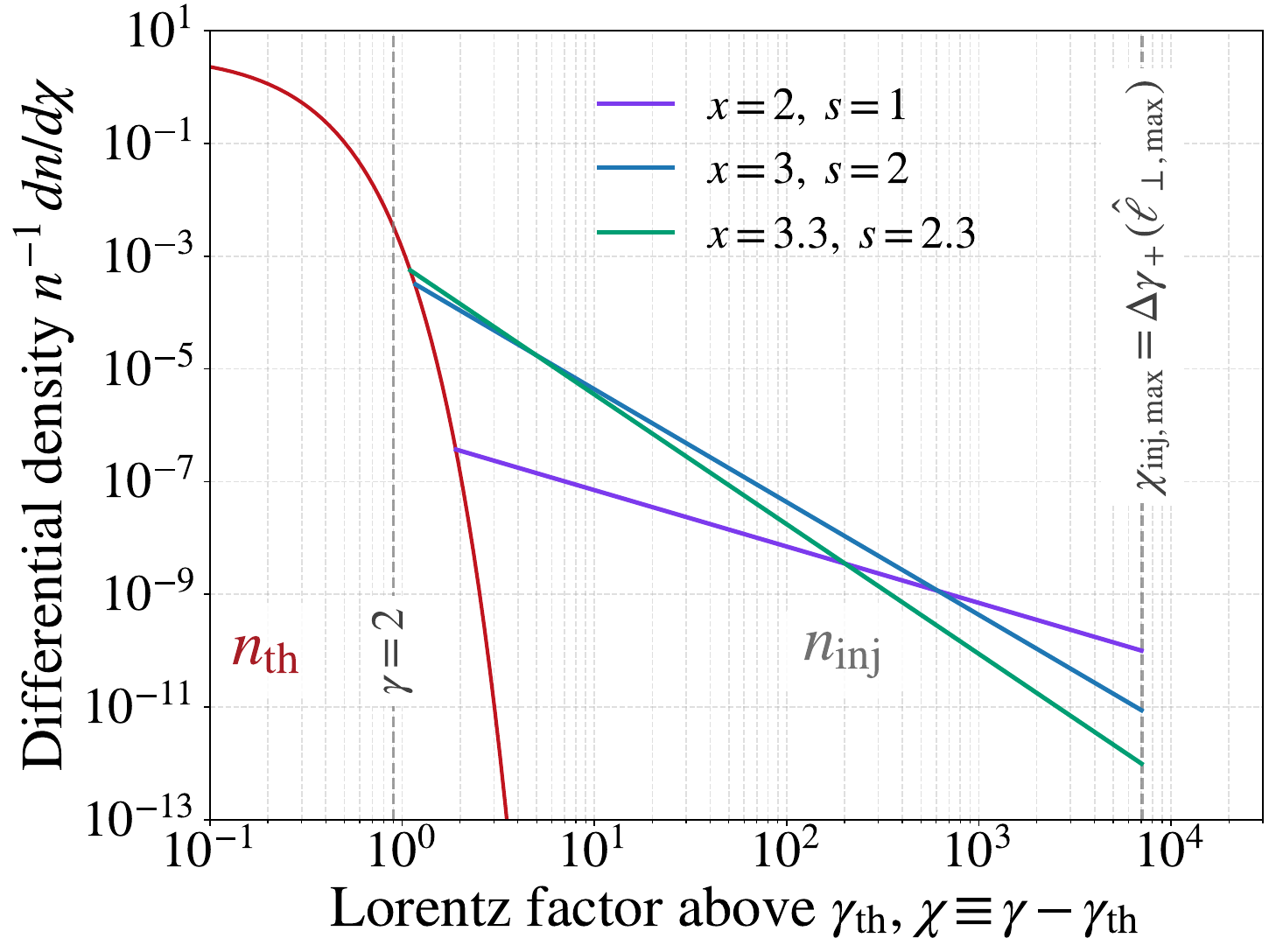}
    \caption{
    Differential densities~$n^{-1}\,dn/d\chi$ of injected particles and of the thermal reservoir vs~$\chi \equiv \gamma - \gamma_{\rm th}$ after one outer turnover time, $t=L_0/(M_A\beta_A c)$. The participating thermal fraction is set to its cascade-power limit~$f_{\rm th}=f_{\rm th}^{\rm max}(x)$ for each power law, so that each source delivers the same energy density,~$\epsilon_{\rm inj}M_A^2B_0^2/8\pi$. The power laws mark the equal-encounter case ($x=2$), the boundary of hard injection ($x=3$), and the MHD value ($x=3.3$). These values span the range suggested by current measurements \citep{Uritsky2010,Zhdankin2016,Serrano2025}. Each power law is drawn from the energy at which it rises above the thermal reservoir (red), a Maxwell--J\"uttner distribution of temperature~$0.1\,mc^2$. This energy lies near the injection threshold~$\chi\approx\gamma_{\rm th}$. Dashed lines mark~$\gamma=2$ and the largest single-encounter gain, $\chi=\Delta\gamma_+(\hat{\ell}_{\perp,\rm max})$. The fiducials used here are $\gamma_{\rm th}=1.1$, $\sigma_c=1$, the cold limit~$\sigma_h=\sigma_c$, $L_0/\mathscr{d}=10^8$, $\ell_{\perp,\rm min}=10\mathscr{d}$, $\hat{\ell}_{\perp,\rm max}=10^{-2}$, $\beta_{\rm rec}=0.1$, $M_A = 0.1$, $\kappa=0$, and $\epsilon_{\rm inj}=1$. Because~$f_{\rm th}^{\rm max}\propto1/p_+$, the plotted curves do not depend on~$p_+$.
    }
    \label{fig:spect}
\end{figure}

The index~$x$ determines which sheets dominate each property of the ensemble. Per logarithmic interval of~$\ell_\perp$, the injected energy scales as~$\ell_\perp^{3+\kappa-x}$, the encounter rate as~$\ell_\perp^{2-x}$, and the dissipated power as~$\ell_\perp^{3+2\kappa-x}$ [Eqs.~\eqref{eq:delta-gamma-traversal}, \eqref{eq:A-cs}, and~\eqref{eq:cs_power}]. The value~$x=2$ spreads the encounters equally over logarithmic intervals and gives~$s=1$. The value~$x=3+\kappa$ spreads the injected energy equally and gives~$s=2$, so the injected spectrum is hard ($s<2$) exactly when the longest sheets carry most of the injected energy. The value~$x=3+2\kappa$ spreads the dissipated power equally and gives~$s=(2+3\kappa)/(1+\kappa)\geq2$, so hard injection requires the longest sheets to dominate the dissipation. For~$\kappa=0$ the last two values coincide at~$x=3$, so hard injection occurs exactly when the longest sheets dominate the dissipation. Intense current sheets in the MHD simulations of~\citet{Zhdankin2016} spread their dissipation nearly uniformly over logarithmic intervals of~$\ell_\perp$, so they lie close to~$x=3+2\kappa$ and near the boundary of hard injection. Kinetic simulations find~$x\approx2$ below about~$30$ skin depths~\citep{Serrano2025}, which would give hard injection, but at the fiducials of Fig.~\ref{fig:spect} only sheets longer than about~$160\mathscr{d}$ lift particles above~$\gamma_{\rm th}$.

Figure~\ref{fig:spect} illustrates Eq.~\eqref{eq:Qinj} at the fiducial $\kappa=0$ for~$x=2$, $3$, and~$3.3$, from equal encounter rates through the boundary of hard injection to the MHD value. The plotted amplitude is fixed by the cascade-power limit of Eq.~\eqref{eq:fth_max}.

%% file: sections/3_applications.tex
\section{Current-sheet statistics and possible applications}
The injection mechanism described in this paper applies to turbulent, low-$\beta$, collisionless plasmas possessing a broad hierarchy of actively reconnecting current sheets, each immersed in a strong guide magnetic field.

The mapping in Eq.~\eqref{eq:s_general_q} does not depend on how the sheet-length index is set. However, how the sheet-length index~$x$ and the other inputs to Eq.~\eqref{eq:s_general_q} depend on environmental parameters and turbulence properties remains poorly understood. A natural next step is therefore to develop and numerically validate a theory of turbulent current-sheet statistics. Ideally, such a theory would be integrated within existing theories of turbulence and of plasmoid formation and statistics \citep{Loureiro2007,Uzdensky2010,Huang2012,LingamComisso2018,Majeski2021}.

Within the measured range~$2\lesssim x\lesssim3.3$, the indices~$2<x<3$ give hard injection spectra~$1<s<2$ for~$\kappa=0$. Several possible applications follow from these spectra, provided the source lies in the strong guide-field, low-$\beta$, collisionless regime described above. A direct phenomenological implication is hard injection of electrons and ions. The reconnection electric field raises only the momentum along the field, and kinetic simulations find that the particles it energizes move preferentially along the field and radiate weakly~\citep{Comisso2019,Comisso2020}. With a strong guide field, energized particles keep their perpendicular momentum, and their pitch angles decrease as their energy grows~\citep{Nattila2022,Vega2025,Humphrey2026a,Humphrey2026b}. If pitch-angle scattering isotropizes the injected particles without changing their energies, and if subsequent effects of acceleration, cooling, or escape do not strongly reshape the slope across an energy range lying within the injection interval, optically thin synchrotron emission gives \(F_\nu\propto\nu^{-\alpha}\), with \(\alpha=(s-1)/2\) \citep{RybickiLightman1979}. An observed optically thin radio index maps to~$x=2+2(1+\kappa)\alpha$, and the boundary of hard injection,~$x=3+\kappa$, corresponds to~$\alpha=1/2$ for any~$\kappa$. The measured range~\(2\lesssim x\lesssim3.3\) then gives flat or mildly steep spectra of~$0\lesssim\alpha\lesssim0.65$ for~$\kappa=0$. The MHD value~$x\simeq3.3$ lies at the steep end of this range, with~$s\simeq2.3$ and~$\alpha\simeq0.65$. Hard injection may help explain flat radio spectra in optically thin sources.

In pulsar wind nebulae, radio-emitting electrons form a hard, long-lived low-energy population not easily produced by a single relativistic shock spectrum \citep{Gaensler2006,Bucciantini2011,Lyutikov2018,Lyutikov2019,Luo2020}. For the Crab Nebula, the radio spectrum has~$\alpha \simeq 0.3$~\citep{Bietenholz1997,Atoyan1999,Lyutikov2019}, which maps to~$s\simeq1.6$ and~$x \simeq 2.6$--$2.8$ (for~$0\leq\kappa\leq1/3$) in this framework, within the hard range~$\alpha<1/2$. A quantitative Crab application, which would require source-specific transport, cooling, escape, and current-sheet statistics, is left to future work.

If the current-sheet hierarchy is present in the jets of active galactic nuclei (AGN), it could supply an intrinsic hard electron component that can coexist with synchrotron self-absorption. In AGN coronae, a current-sheet ensemble may inject ions into subsequent stochastic acceleration \citep{Lemoine2025} and provide a seed population for the hard low-energy proton component invoked in models of neutrino emission from NGC~1068 \citep{Murase2020b,Fiorillo2024,Mbarek2024,Fiorillo2024b,Karavola2025a,Karavola2026}.

%% file: sections/4_end.tex
\section{Multi-sheet drift and diffusion}
Here I calculate the energy-space drift and diffusion of particles as they traverse the current-sheet ensemble. As in Eq.~\eqref{eq:encounter-probability}, a particle of Lorentz factor~$\gamma$ interacts only with sheets longer than~$\ell_{\perp,\gamma}=\max(\ell_{\perp,\rm min},\gamma\rho_0)$. The accessible current-sheet encounter rate is therefore
\begin{equation} \label{eq:nu-gamma}
\begin{split}
    \nu_\gamma
    &= \frac{c}{V}\int_{\hat{\ell}_{\perp,\gamma}}^{\hat{\ell}_{\perp,\rm max}} A_{\rm cs} \frac{dN_{\rm cs}}{d\hat{\ell}_\perp}\,d\hat{\ell}_\perp \\
    &= M_A \, \beta_{\rm rec}^{-1} \, \frac{c\,\mathscr{d}}{L_0^2} \frac{\mathcal{I}_{2-x}(\hat{\ell}_{\perp,\gamma},\hat{\ell}_{\perp,\rm max})}{\mathcal{I}_{3+2\kappa-x}(\hat{\ell}_{\perp,\rm min},\hat{\ell}_{\perp,\rm max})},
\end{split}
\end{equation}
where the second equality uses~$A_{\rm cs} \sim 2\mathscr{d}\ell_\perp$ and the normalization of Eq.~\eqref{eq:sheet-pop-normalization}. Once in a sheet, a particle resides there for the traversal time~$\ell_\parallel/c$, whose encounter-weighted mean is
\begin{equation} \label{eq:ell-parallel-gamma}
\begin{split}
    \frac{\ell_{\parallel, \gamma}}{c}
    &\equiv \int_{\hat{\ell}_{\perp, \gamma}}^{\hat{\ell}_{\perp,\rm max}} \mathcal P_\gamma \frac{\ell_\parallel}{c} \, d\hat{\ell}_\perp \\
    &= M_A^{-1} \, \frac{L_0}{c} \, \frac{\mathcal{I}_{3-\kappa-x}(\hat{\ell}_{\perp,\gamma},\hat{\ell}_{\perp,\rm max})}{\mathcal{I}_{2-x}(\hat{\ell}_{\perp,\gamma},\hat{\ell}_{\perp,\rm max})}.
\end{split}
\end{equation}
The current-sheet encounter rate that accounts for finite residence times is thus
\begin{equation} \label{eq:nu-ms}
    \nu_\gamma^{\rm ms} \equiv \frac{1}{1/\nu_\gamma + \ell_{\parallel,\gamma}/c},
\end{equation}
where~$\nu_\gamma$ is given by Eq.~\eqref{eq:nu-gamma} and~$\ell_{\parallel,\gamma}$ is given by Eq.~\eqref{eq:ell-parallel-gamma}.

Meanwhile, averaging over the accessible current sheets gives the moments of the positive energy jump of a particle of Lorentz factor~$\gamma$,
\begin{equation} \label{eq:jump-moments}
\begin{split}
    \mu_j(\gamma) &\equiv
    \int_{\hat{\ell}_{\perp, \gamma}}^{\hat{\ell}_{\perp,\rm max}} \mathcal P_\gamma \, [\Delta\gamma_+(\hat{\ell}_\perp)]^j \,d\hat{\ell}_\perp \\
    &= (\Delta\gamma_0)^j \, \frac{\mathcal{I}_{2+j(1+\kappa)-x}(\hat{\ell}_{\perp,\gamma},\hat{\ell}_{\perp,\rm max})} {\mathcal{I}_{2-x}(\hat{\ell}_{\perp,\gamma},\hat{\ell}_{\perp,\rm max})},
\end{split}
\end{equation}
so that~$\mu_1$ is the mean positive jump. Assuming that positive and negative work events have statistically symmetric magnitudes, the jump moments of Eq.~\eqref{eq:jump-moments} and the encounter rate~$\nu_\gamma^{\rm ms}$ of Eq.~\eqref{eq:nu-ms}, which includes residence times, give the Kramers--Moyal \citep{Kramers1940,Moyal1949} coefficients
\begin{equation} \label{eq:KM-coeffs}
    \mathcal{A}_\gamma = (2p_+-1) \, \nu_\gamma^{\rm ms} \, \mu_1,
    \qquad
    \mathcal{D}_{\gamma\gamma} = \frac{1}{2} \, \nu_\gamma^{\rm ms} \, \mu_2 .
\end{equation}
Thus \(p_+=1/2\) gives no systematic drift but leaves finite energy-space diffusion. For trajectories biased toward positive-work encounters, $p_+>1/2$, the multi-sheet drift time~$\tau_\gamma^{\rm ms} = \gamma/\mathcal{A}_\gamma$ evaluates to
\begin{equation} \label{eq:tau-ms}
\begin{split}
    \omega_c \tau^{\rm ms}_\gamma
    &=
    \frac{\gamma\,\beta_A^{-1}\beta_{\rm rec}^{-1} M_A^{-2}}{2p_+-1}
    \frac{\mathcal{I}_{3-\kappa-x}(\hat{\ell}_{\perp,\gamma},\hat{\ell}_{\perp,\rm max})}{\mathcal{I}_{3+\kappa-x}(\hat{\ell}_{\perp,\gamma},\hat{\ell}_{\perp,\rm max})}
    \\
    &\times \left[
    1+
    \beta_{\rm rec}\frac{L_0}{\mathscr{d}}
    \frac{\mathcal{I}_{3+2\kappa-x}(\hat{\ell}_{\perp,\rm min},\hat{\ell}_{\perp,\rm max})}{
    \mathcal{I}_{3-\kappa-x}(\hat{\ell}_{\perp,\gamma},\hat{\ell}_{\perp,\rm max})}
    \right].
\end{split}
\end{equation}

\begin{figure}[tp!]
    \centering
    \includegraphics[width=\linewidth]{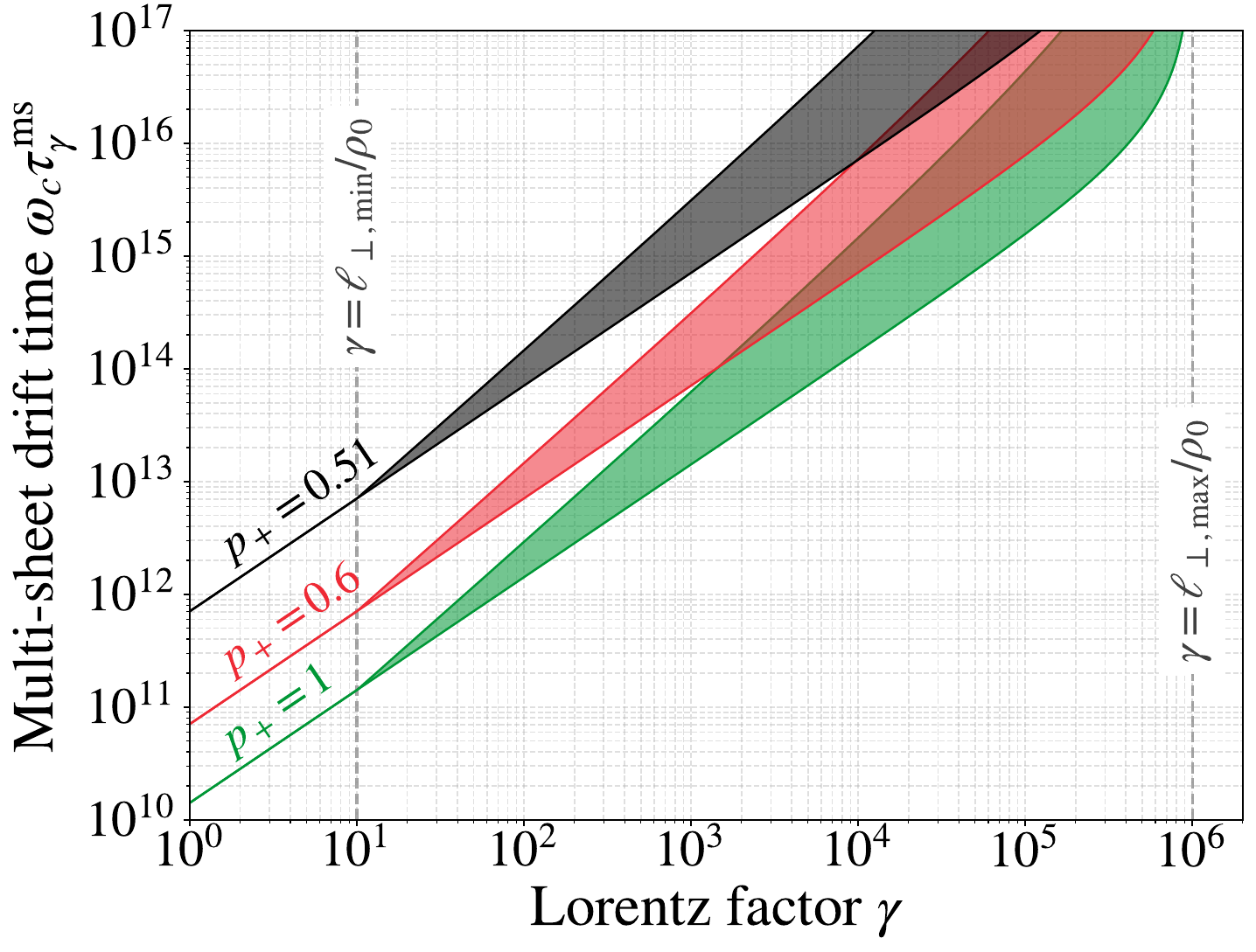}
    \caption{
    Multi-sheet drift time from Eq.~\eqref{eq:tau-ms} evaluated at~$M_A = 0.1$, $\kappa=0$, \(L_0/\mathscr{d}=10^8\), $\beta_{\rm rec}=0.1$, and \(\sigma_c=1\), using the cold limit \(\sigma_h=\sigma_c\), so that \(\beta_A=[\sigma_c/(1+\sigma_c)]^{1/2}\). Shaded bands show $x\in[2,\,3.3]$ for~$p_+=0.51$ (black), $0.6$ (red), and $1$ (green). For~$\kappa=0$ the drift time does not depend on~$x$ below~$\gamma=\ell_{\perp,\rm min}/\rho_0$, where the bands reduce to lines. No positive multi-sheet drift time is defined for \(p_+\leq1/2\), although energy-space diffusion remains finite. Dashed vertical lines mark \(\gamma=\ell_{\perp,\rm min}/\rho_0\) and \(\gamma=\ell_{\perp,\rm max}/\rho_0\), with~$\ell_{\perp,\rm min}=10\mathscr{d}$ and~$\hat{\ell}_{\perp,\rm max}=10^{-2}$.
    }
    \label{fig:tau-acc}
\end{figure}

Figure~\ref{fig:tau-acc} shows the resulting multi-sheet drift time for~$x \in [2,\,3.3]$. The drift time lengthens as~$p_+ \to 1/2$, where the systematic drift vanishes and only energy-space diffusion survives. It also diverges as~$\gamma$ approaches~$\ell_{\perp,\rm max}/\rho_0$, above which the gyroradius of a particle exceeds the longest sheet in the ensemble, so the particle no longer takes part in multi-sheet acceleration. In a realistic system, the high-energy cutoff may instead be set by cooling or escape. The coefficients of Eq.~\eqref{eq:KM-coeffs} assume jumps small compared with~$\gamma$, i.e., $\gamma\gg\mu_2/\mu_1$, which for the parameters of Fig.~\ref{fig:tau-acc} holds only for~$\gamma\gtrsim10^4$, so at lower~$\gamma$ the drift time of Fig.~\ref{fig:tau-acc} is an order-of-magnitude estimate.